\documentclass[aps,prb,showpacs,twocolumn,amsmath,amssymb,superscriptaddress,letterpaper]{revtex4}
\usepackage{times}
\usepackage{amsfonts}
\usepackage{mathrsfs}
\usepackage{graphicx}
\usepackage{float}
\usepackage{dcolumn}
\usepackage{bm}
\usepackage{color}

\usepackage[colorlinks,bookmarks=false,citecolor=blue,linkcolor=red,urlcolor=blue]{hyperref}
\usepackage{multirow}

\def\be{\begin{equation}}       \def\ee{\end{equation}}
\def\bea{\begin{eqnarray}}      \def\eea{\end{eqnarray}}

\begin{document}
\title{  Topological Critical Materials of Ternary Compounds }

\author{Shengshan Qin}
\affiliation{Beijing National Laboratory for Condensed Matter Physics,
and Institute of Physics, Chinese Academy of Sciences, Beijing 100190, China}
\affiliation{University of Chinese Academy of Science, Beijing 100049, China}

\author{Congcong Le}
\affiliation{Kavli Institute of Theoretical Sciences, University of Chinese Academy of Sciences,
Beijing, 100049, China}
\affiliation{Beijing National Laboratory for Condensed Matter Physics,
and Institute of Physics, Chinese Academy of Sciences, Beijing 100190, China}

\author{Xianxin Wu}
\affiliation{Institute for Theoretical Physics and Astrophysics,
Julius-Maximilians University of W“urzburg, Am Hubland, D-97074 W“urzburg, Germany}

\author{Jiangping Hu}\email{jphu@iphy.ac.cn}
\affiliation{Beijing National Laboratory for Condensed Matter Physics,
and Institute of Physics, Chinese Academy of Sciences, Beijing 100190, China}
\affiliation{Kavli Institute of Theoretical Sciences, University of Chinese Academy of Sciences,
Beijing, 100049, China}
\affiliation{Collaborative Innovation Center of Quantum Matter,
Beijing 100190, China}

\date{\today}

\begin{abstract}
 We review topological properties of two series of ternary compounds  AMgBi (A=K, RB, Cs) and ABC with a hexagonal ZrBeSi type structure. The first series of materials AMgBi are   predicted to be topological critical Dirac semimetals. The second series of ternary compounds, such as KZnP, BaAgAs, NaAuTe and KHgSb, can be used to realize various topological insulating states and semimetal states. The states are highly tunable as the realization of these topological states depends on the competition between several energy scales, including the energy of atomic orbitals, the energy of crystal splitting, the energy difference between  bonding and antibonding states, and the strength of  spin-orbit coupling. The exotic surface states in these series of compounds are predicted and are closely related to their unique crystal structures.
\end{abstract}


\maketitle

\section{Introduction}

Pioneered by the discovery of integer quantum Hall effect\cite{IQHE} (IQHE), the research of new topological states in materials has made great progress both theoretically and experimentally in the past decades. Materials with nontrivial topology are characterized by their exotic edge (surface) states. According to the bulk-edge correspondence  principle, the appearance of the edge or surface states is protected by the nontrivial topology in  the bulk band structures\cite{TKNN}.

For a long time, the research of topological states has been mainly focused on  insulating states\cite{TI_review1,TI_review2,TI_review3,TI_review4}. The quantum spin Hall (QSH) state\cite{2DTI_Kane1,2DTI_Kane2,2DTI_Kane3,2DTI_Zhang}, i.e., the two-dimensional (2D) topological insulator (TI) state, first proposed in graphene, has  greatly accelerated the research of topological materials. The QSH state is protected by  time reversal symmetry (TRS). The bulk bands of the 2D TIs carry a topological nontrivial $Z_2$ number. On the edges, the topological nontrivial  $Z_2=1$ state is characterized by a  robust Dirac cone. Soon the concept of  the topological insulator was generalized to three-dimensional (3D) systems\cite{3DTI}. Spin-orbit coupling (SOC) has played an essential role in the realization of  TIs. Compared to the ordinary insulators, in TIs, the conduction band bottom and the valance band top are inverted due to the strong SOC \cite{BHZ}.

Besides the insulating state, the research of topological phases has also been extended to the metallic states of materials. The most well known topological metallic phase is topological semimetal\cite{TI_review3,TI_review4,WS review} (TS) phase including topological Weyl semimetals\cite{HgCrSe,XG Wang,multilayer weyl,nodal Balents,TRS_WS Balents,TaAs thory1,TaAs thory2} (TWSs) and topological Dirac semimetals\cite{Dirac Kane,Na3Bi,Cr3As2} (TDSs). The TS phase can be viewed as an intermediate state of the TI phase and the normal insulator (NI) phase\cite{intermediate1,intermediate2} in general.  The low energy excitations in the band structures of a TWS obey the Weyl equation.  The Weyl points can be viewed as monopoles\cite{monopole} and    always appear in pairs\cite{Weyl pair1,Weyl pair2}.   On the surfaces, the TWSs are characterized by Fermi arcs\cite{XG Wang}.   Similar to the TWSs, the surface states of a TDS  also exhibit Fermi arcs\cite{Na3Bi,Cr3As2}.

Besides time reversal symmetry, one can also use other discrete symmetries to classify topological states, such as   topological crystalline insulator (TCI) proposed by L. Fu\cite{TCI}. The states have been realized in SnTe family materials\cite{SnTe1,SnTe2}.  The classification of the TSs has also been extended.  For example, they can be classified into type-\uppercase\expandafter{\romannumeral1} and type-\uppercase\expandafter{\romannumeral2} according to the dispersion of their bands\cite{type2}, and the type-\uppercase\expandafter{\romannumeral2} semimetal phase has been realized in PdTe$_2$ and PtTe$_2$\cite{PdTe1,PdTe2,PtTe}.   Furthermore,  new fermions which have no correspondence in particle physics have been predicted \cite{new fermion}, and a three-component fermion has been observed in MoP\cite{MoP}, recently.

To explore topological physics and  its promising future applications,  finding more materials with nontrivial topology  is of great significance. In this paper, we mainly review two series of ternary compounds, where various topological phases can be realized. We first review some theoretical concepts of topological orders. Then  we discuss a series of materials AMgBi (A$=$K, Rb, Cs), which are predicted to be the topological critical Dirac semimetals and  a series of ternary compounds KZnP, BaAgAs, NaAuTe and KHgSb, which belong to the space group $P6_3/mmc$, in which various topological phases could be realized. The mechanism leading to different topological phases are discussed in detail. Finally we  briefly summarize our results and discuss future researches in these series of materials.

\begin{figure}
\centerline{\includegraphics[width=0.45\textwidth]{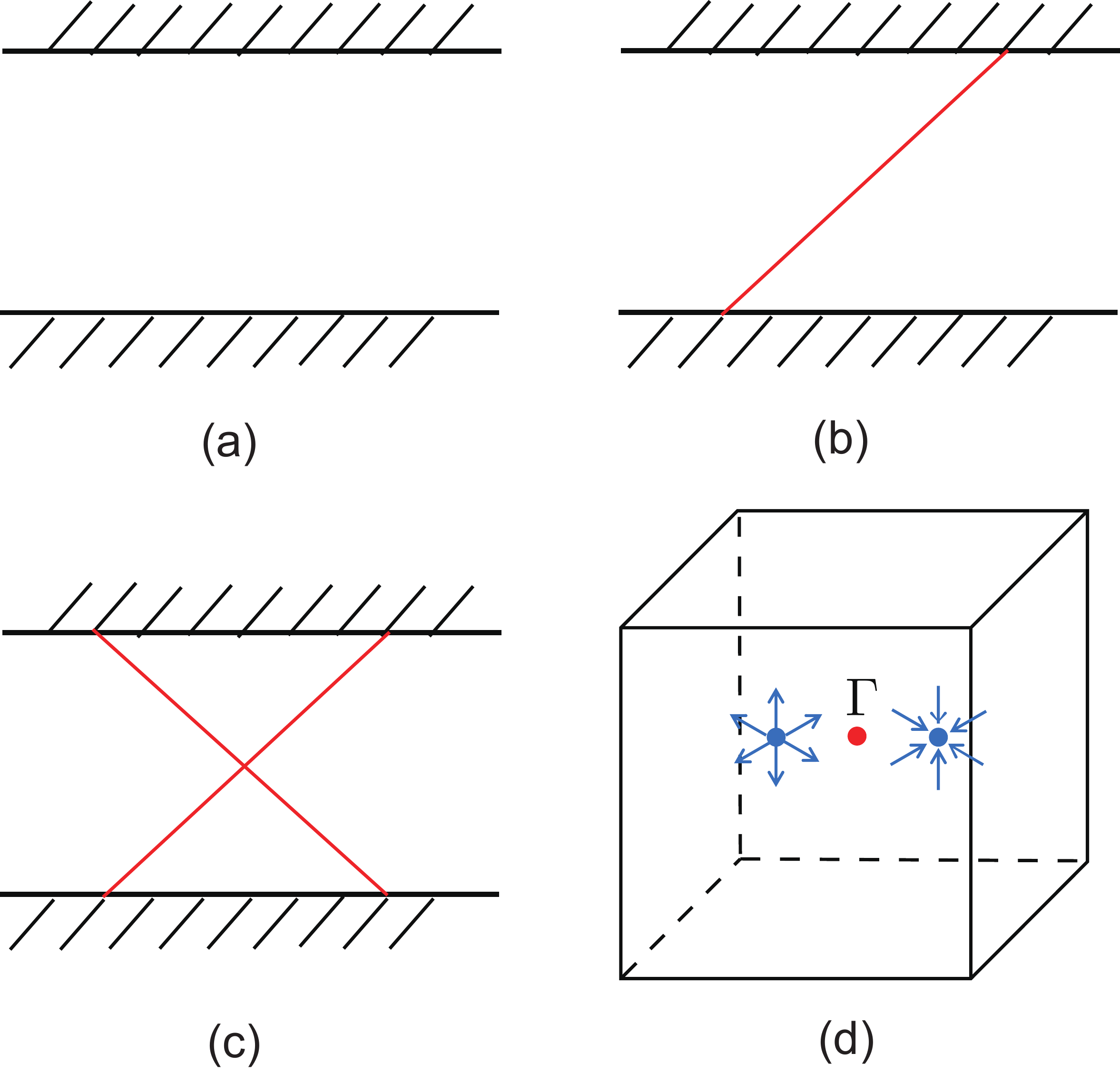}}
\caption{(color online) Illustrative figures for NIs, TIs and TWSs. (a) (b) and (c) show the edge states for a NI, a Chern insulator and a TI, respectively. For a NI, there are no gapless states on its edge, and there are $n$ gapless chiral modes for a Chern insulator with Chern number $n$. For a TI protected by the TRS, a robust Dirac cone appears on its edge, which is contributed by the Kramers' pairs. The Weyl points in the bulk bands can be viewed as 'magnetic monopoles', which are source or sink of the Berry curvature, as shown in (d).
\label{AMgBicrystal}}
\end{figure}

\section{Basic Concept of Topological Invariants}

The topological materials are classified by topological nontrivial invariants. In the following, we simply review the concept of the $Z_2$ topological invariant for TIs, the topological invariant for the TCIs  protected by mirror symmetries and  the topology of the TSs.

The $Z_2$ topological index is proposed by Kane and Mele to classify insulators into TIs and NIs\cite{2DTI_Kane1} for the materials with TRS. There are several different ways to get the same $Z_2$ topological invariant. Here, we introduce the $Z_2$ topological invariant in the point of view of the evolution of the charge center of the Wannier states\cite{spin pump}. In this definition, the $Z_2$ topological invariant can be understood according to the time reversal polarization for all the occupied states

\begin{eqnarray}\label{Z2_define}
(-1)^\nu &=& P_T(k_x=\pi)-P_T(k_x=0), \\
P_T(k_x) &=& P^{\uppercase\expandafter{\romannumeral1}}(k_x)-P^{\uppercase\expandafter{\romannumeral2}}(k_x), \nonumber \\
P^s(k_x) &=& \frac{1}{2\pi}\int^\pi_{-\pi}dk_y A^s(k_x,k_y), \nonumber \\
A^s(k_x,k_y) &=& i\sum_a\langle u^s_a(k_x,k_y)|\partial_{k_y}|u^s_a(k_x,k_y)\rangle, s={\uppercase\expandafter{\romannumeral1}},{\uppercase\expandafter{\romannumeral2}} \nonumber
\end{eqnarray}
where $|u^{{\uppercase\expandafter{\romannumeral1}}}_a(k_x,k_y)\rangle$ and $|u^{{\uppercase\expandafter{\romannumeral2}}}_a(k_x,k_y)\rangle$ are Kramers' pairs, $a$ denotes the non-Kramers' pair indices for all of the occupied bands, and $A^s(k_x,k_y)$ is the well known Berry connection. After some algebra, Eq.\ref{Z2_define} comes to the form

\begin{eqnarray}\label{Z2_final}
(-1)^\nu &=& \prod_i \frac{\sqrt{det[B({\bf \Gamma_i})]}}{pf[B({\bf \Gamma_i})]}, i=1,2,3,4 \\
B_{mn}^{st}({\bf k}) &=& \langle u^s_m(-{\bf k})|T|u^t_n({\bf k})\rangle, \nonumber
\end{eqnarray}
here, ${\bf \Gamma_i}$ label the four time reversal invariant points in the first Brillouin zone (BZ), $T$ is the TRS operator, and $pf[]$ denotes the Pfaffian of the skew-symmetric matrix in the bracket. If the 2D system have inversion symmetry (IS), Eq. \ref{Z2_final} will have a much simpler form\cite{Z2 inversion}

\begin{eqnarray}\label{Z2_inversion}
(-1)^\nu &=& \prod_{i,a}\xi^{{\uppercase\expandafter{\romannumeral1}}}_a({\bf \Gamma_i}), \\
P|u^{{\uppercase\expandafter{\romannumeral1}}}_a({\bf \Gamma_i})\rangle &=& \xi^{{\uppercase\expandafter{\romannumeral1}}}_a({\bf \Gamma_i})|u^{{\uppercase\expandafter{\romannumeral1}}}_a({\bf \Gamma_i})\rangle, \nonumber
\end{eqnarray}
where $P$ is the IS operator and $\xi^{{\uppercase\expandafter{\romannumeral1}}}_a({\bf \Gamma_i})$ is the parity of state $|u^{{\uppercase\expandafter{\romannumeral1}}}_a({\bf \Gamma_i})\rangle$ at the time reversal invariant point ${\bf \Gamma_i}$. The Kramers' pairs satisfy $|u^{{\uppercase\expandafter{\romannumeral1}}}_a(-k_x,-k_y)\rangle=T|u^{{\uppercase\expandafter{\romannumeral2}}}_a(k_x,k_y)\rangle$ up to a $U(1)$ gauge. Since the point group symmetries commute with the TRS, two states of a Kramers' pair always have the same parity. Thus, only the state $|u^{{\uppercase\expandafter{\romannumeral1}}}_a({\bf \Gamma_i})\rangle$ for each Kramers' pair is taken in Eq.\ref{Z2_inversion}.

The 3D TIs are also protected by the TRS, and there are four topological invariants $\nu_0;(\nu_1\nu_2\nu_3)$ to characterize their topological states\cite{3DTI}. The three topological invariants $\nu_{1,2,3}$, similar to the 2D $Z_2$ topological number defined above, describe the topological states in the three independent time reversal invariant 2D planes. The form of $\nu_0$ is given by

\begin{eqnarray}
(-1)^{\nu_0} &=& \prod_i \frac{\sqrt{det[B({\bf \Gamma_i})]}}{pf[B({\bf \Gamma_i})]},i=1,2,3,\cdots,8
\end{eqnarray}
where $B({\bf \Gamma_i})$ is defined the same as in Eq.\ref{Z2_final}, and ${\bf \Gamma_i}$ are the eight time reversal invariant points in the 3D BZ. The four topological invariants $\nu_0;(\nu_1\nu_2\nu_3)$ clearly classify the 3D insulating states into three different topological states: strong topological insulator (STI), weak topological insulator (WTI) and NI.

The TCI  states are protected by crystalline symmetries\cite{TCI}.  Here we review  the topological invariants of the  TCIs that are classified  by mirror symmetry, which have been observed in SnTe family materials\cite{SnTe2}.  For a 3D system with a mirror symmetry, there must be mirror invariant planes in  3D BZ. For instance, considering a mirror reflection $M_z$ perpendicular to the $z$-direction, there are two $M_z$ invariant planes:  $(k_x,k_y,0)$ and $(k_x,k_y,\pi)$.  The mirror symmetry can be viewed as a combination of the IS and a two-fold rotational symmetry.  Specifically, $M_z=PC_{2z}$ where P is the IS  and $C_{2z}$ is the two-fold rotational symmetry along the $z$-axis. Thus, in the mirror invariant planes for a spinful system, the wavefunctions must satisfy $M_z|\varphi_\eta({\bf k})\rangle=\eta|\varphi_\eta({\bf k})\rangle$ where $\eta=\pm i$ are the eigenvalues of $M_z$. As a result, $|\varphi_\eta({\bf k})\rangle$ can be classified into two different subspaces according to its mirror eigenvalue $\eta$. In each subspace of $M_z$, there is a Chern number $N_\eta$. The total Chern number $N$ for the whole system can be expressed as the summation of the Chern number in all of the subspaces: $N=N_{+i}+N_{-i}$. However, besides the total Chern number, a new topological invariant, so-called mirror Chern number\cite{SnTe1,TCI review}, can also be well-defined: $N_M=(N_{+i}-N_{-i})/2$.   The  mirror Chern number can be nonzero when the the total Chern number for a system is zero. The TCIs characterized by the mirror Chern number are protected by the mirror symmetry.  The surface states of this kind of TCIs   strongly depend  on the symmetry and direction of   surfaces. For a TCI  with the mirror Chern number  $N_M=n$, it has $n$ Dirac cones on its surface, where the mirror symmetry is preserved.

For topological semimetals, the Fermi surfaces in the bulk are classified by the appearance of Weyl/Dirac points.  In 3D systems, the Weyl point has linear dispersion in all directions and behaves as a magnetic monopole\cite{monopole}, which is a source or sink of the Berry curvature in the bulk band structures. The topology of the Weyl point is characterized by its 'magnetic charge', the chirality of the Weyl point $\chi$\cite{WS review,XG Wang}, which is equal to the Chern number carried by the FS enclosing the Weyl point, $\chi=\frac{1}{2\pi}\oint_{FS}\Omega({\bf k})\cdot dS({\bf k})$, where $\Omega({\bf k})$ is the Berry curvature on the FS. Furthermore, the low energy dispersion of a Weyl point can be described by a minimal two bands model, which has a general form $h({\bf k})=k_iA_{ij}\sigma_j$, where $\sigma_j$ are the three Pauli matrixes and $A_{ij}$ are real numbers. The chirality of the Weyl point has another form $\chi=sgn[det(A)]$. Based on the definition of the chirality $\chi$, it is easy to get the chirality of the Weyl points which are connected with each other by some symmetries. For example, the Weyl points relating each other by mirror symmetry and IS have opposite chirality, while the rotational symmetry and TRS preserve their chirality, for the 3D TWSs.

Recently, it is found that the TWS can be classified into type-\uppercase\expandafter{\romannumeral1} and type-\uppercase\expandafter{\romannumeral2} according to its dispersion in the band structures\cite{type2}. As shown in Ref.\cite{type2}, the most general effective low energy model for a Weyl point takes the form

\begin{eqnarray}\label{type2DS}
H({\bf k})=\sum_{i=1,2,3}k_iA_{i0}\sigma_0+\sum_{i,j=1,2,3}k_iA_{ij}\sigma_j,
\end{eqnarray}
where $\sigma_0$ is the two dimensional unit matrix. The energy spectrum for Eq.\ref{type2DS} is as follows

\begin{eqnarray}\label{type2band}
E_\pm({\bf k})&=&\sum_{i=1,2,3}k_iA_{i0}\pm\sqrt{\sum_{j=1,2,3}(\sum_{i=1,2,3}k_iA_{ij})^2}\nonumber\\
              &=&T({\bf k})\pm U({\bf k}),
\end{eqnarray}
According to the band dispersion in Eq.\ref{type2band}, TWSs can be classified into two types: for type-\uppercase\expandafter{\romannumeral2} TWSs, there exists at least one direction in the reciprocal space for which  $|T({\bf k})|>|U({\bf k})|$; Otherwise, it is of type-\uppercase\expandafter{\romannumeral1}. It is easier to understand the difference between the two TWS phases from the point of view of their FSs: for a perfect type-\uppercase\expandafter{\romannumeral1} TWS, its FSs are several discrete points; while for a perfect type-\uppercase\expandafter{\romannumeral2} TWS, its FSs contain electron pockets and hole pockets which connect with each other at the Weyl points.

TDSs can be viewed as special cases of TWSs. A massless Dirac point is four-fold degenerate and it can be viewed as two Weyl points with opposite chirality. When two Weyl points with opposite chirality come together, both of the Weyl points will gap out\cite{Dirac Kane}. Consequently, for most cases, but not for all the cases\cite{Cr3As2}, both TRS and IS are necessary for a TDS, and extra crystalline symmetries are also needed to stabilize the Dirac points\cite{Dirac Kane,DS Nagaosa,DS nonsymmorphic}. The surface states of a TDS are also characterized by Fermi arcs, but the condition is more complicated than the case of a TWS\cite{DS pnas}. Depending on the band dispersion, TDSs can be classified into type-\uppercase\expandafter{\romannumeral1} and type-\uppercase\expandafter{\romannumeral2} similarly\cite{KMgBi,VAl}. Furthermore, TWS phases can usually be realized, if the TRS or IS is broken in a TDS\cite{Dirac Kane,Na3Bi}.

In the following two sections, the topological phases in two series of ternary compounds are analyzed. The calculations of the band structures are performed using density functional theory (DFT) as implemented in the Vienna ab initio simulation package (VASP) code \cite{Kresse1993,Kresse1996,Kresse1996B}. The Perdew-Burke-Ernzerhof (PBE) exchange-correlation functional and the projector-augmented-wave (PAW) approach are used. Throughout the work, the cutoff energy is set to be 500 eV for expanding the wave functions into plane-wave basis. In the calculation, the BZ is sampled in the k space within Monkhorst-Pack scheme\cite{MonkhorstPack}. On the basis of the equilibrium structure, the k mesh used is $10\times10\times6$. We relax the lattice constants and internal atomic positions with GGA, where the plane wave cutoff energy is 600 eV. Forces are minimized to less than 0.01 eV/\AA~ in the relaxation.

\section{topological critical Dirac semimetal AMgBi}

In this section, we review a series of topological critical Dirac semimetals AMgBi (A$=$K, Rb, Cs) whose space group is $P4/nmm$. It is found that AMgBi (A$=$K, Rb, Cs) are symmetry protected Dirac semimetals located near the boundary of type-\uppercase\expandafter{\romannumeral1} and type-\uppercase\expandafter{\romannumeral2} Dirac semimetal phases\cite{KMgBi}. The transition between the two topological phases can be driven by in-plane compressive strain or by doping Rb or Cs into KMgBi. In the following, we discuss KMgBi and RbMgBi in detail. The results for CsMgBi are similar to those of RbMgBi.

\begin{figure}
\centerline{\includegraphics[width=0.45\textwidth]{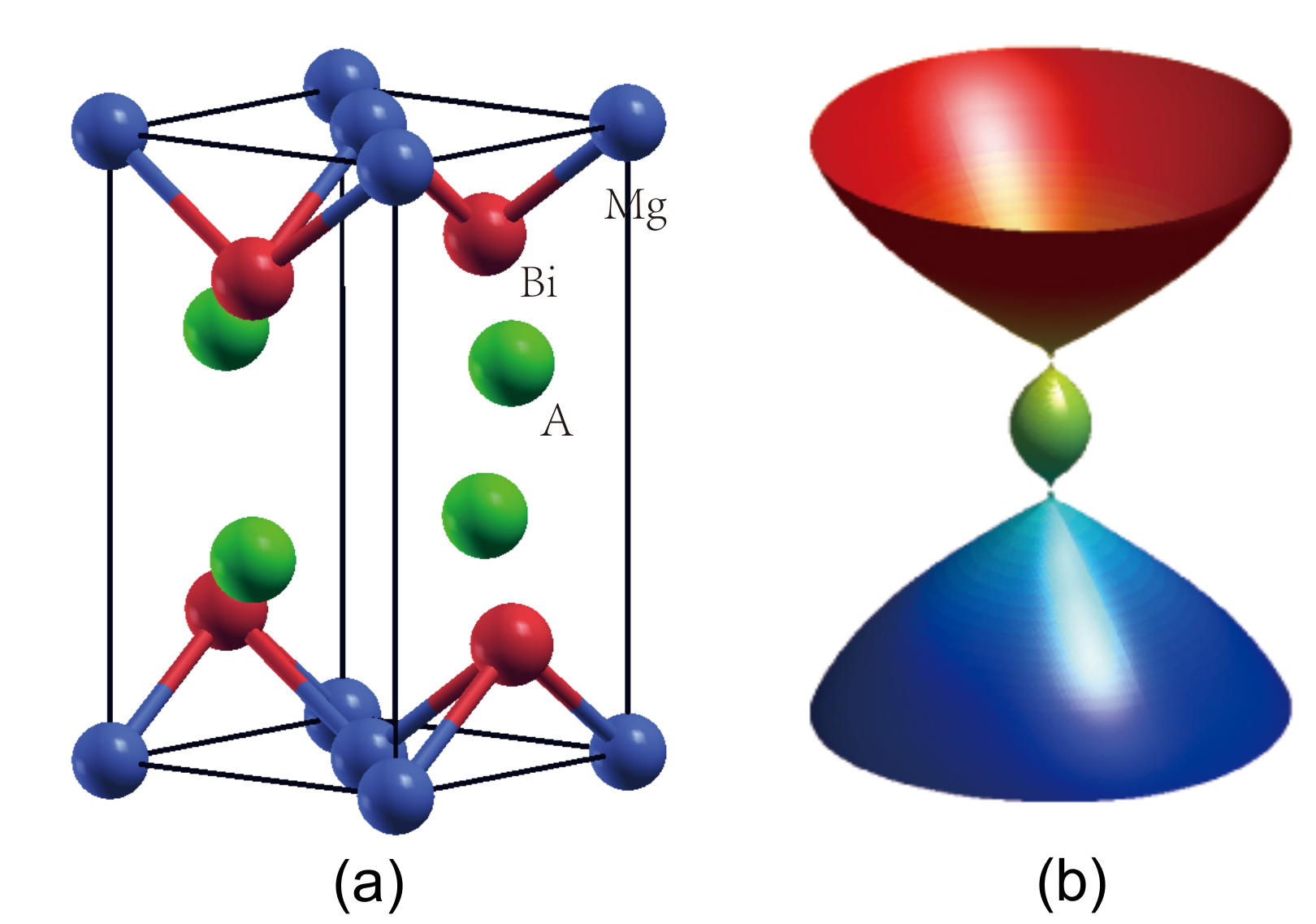}}
\caption{(color online) (a) Crystal structure for AMgBi (A$=$K, Rb, Cs). (b)  the FSs of RbMgBi by lowering Fermi level  $37.5$ meV.
\label{AMgBicrystal}}
\end{figure}

The crystal structure of AMgBi (A$=$K, Rb, Cs)\cite{Rainer1979} with the space group $P4/nmm$ is shown in Fig.\ref{AMgBicrystal}. Its crystal structure is similar to the 111 family of iron-based superconductors\cite{Michael2008,Joshua2008,Dinah2009}. The Mg$_2$Bi$_2$ layers possess an anti-PbO-type atom arrangement, consisting of a square lattice sheet of Mg coordinated by Bi above and below the plane to form face sharing MgBi$_4$ tetrahedra. The bond angle of MgBi$_4$ tetrahedra is very close to that of the perfect tetrahedron.

\begin{figure}
\centerline{\includegraphics[width=0.45\textwidth]{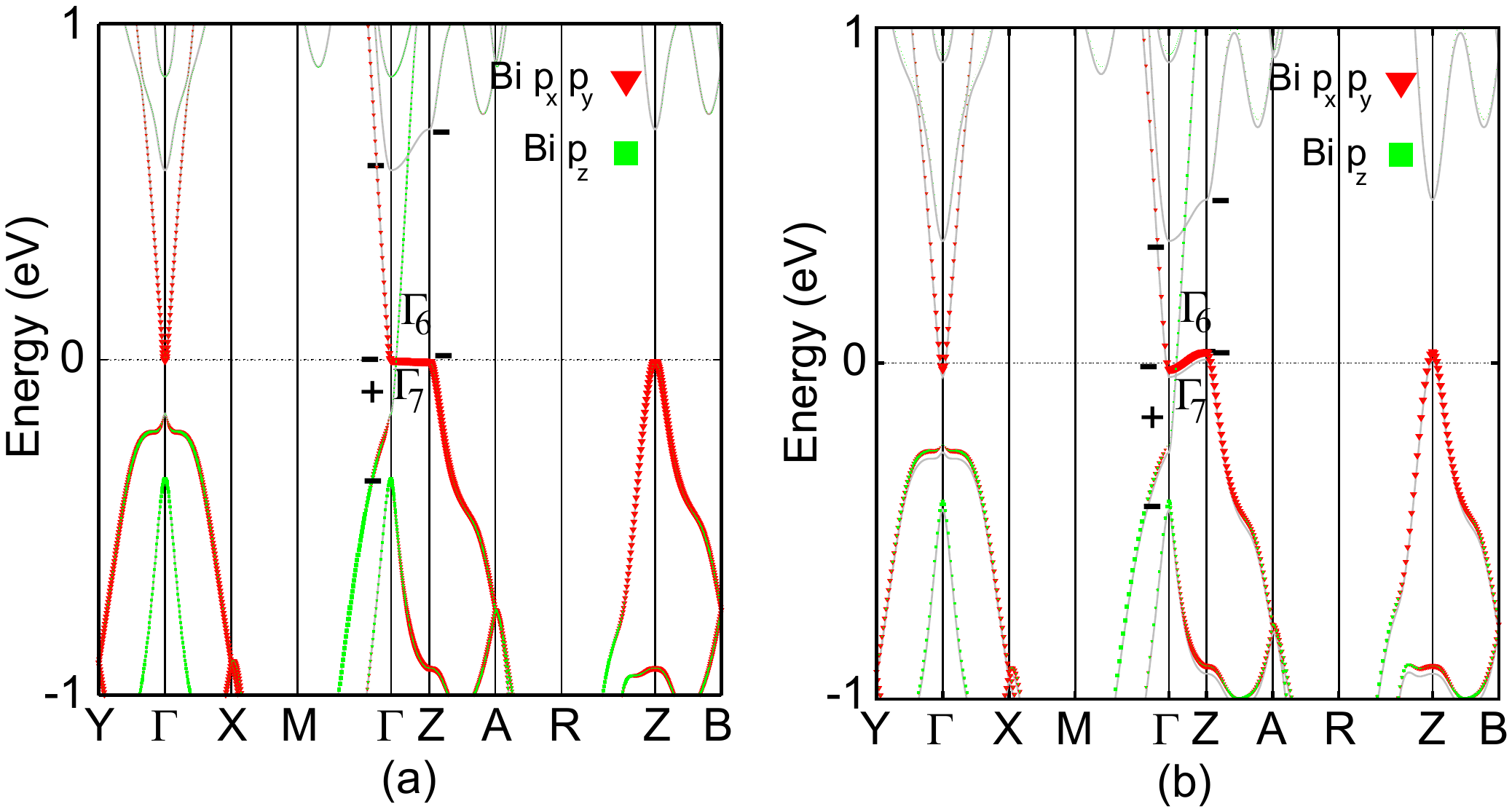}}
\caption{(color online) Band structures for KMgBi (a) and RbMgBi (b)  with SOC. For both cases, the $\Gamma_7$ band with odd parity and $\Gamma_6$ band with even parity are inverted in the $k_z=0$ plane. These two bands have two robust band crossings at ($0, 0,\pm k_z^c$) to form two Dirac points.
\label{AMgBiband}}
\end{figure}

The band structures of KMgBi and RbMgBi are displayed in Fig.\ref{AMgBiband}. Without SOC, both KMgBi and RbMgBi are narrow-gap semiconductors. With the large SOC  in Bi atoms, a band inversion occurs around the $\Gamma$ point between the $\Gamma_7$ and the $\Gamma_6$ bands, which belong to two different irreducible representations, shown in Fig.\ref{AMgBiband}. Near the $\Gamma$ point, the valence band $\Gamma_6$ is mainly attributed to the Bi-6p$_z$ orbital, while the conduction band $\Gamma_7$ is mainly attributed to the Bi-6p$_{x,y}$ orbitals. Furthermore, the $z$-component of the angular momentum of the $\Gamma_7$ band is $J_z=\frac{3}{2}$, while $J_z=\frac{1}{2}$ for the $\Gamma_6$ band. Along the $z$-direction, the $\Gamma_7$ and the $\Gamma_6$ bands have two band crossings at ($0, 0,\pm k_z^c$), which form two Dirac points. The two Dirac points are robust, because the $\Gamma_7$ and the $\Gamma_6$ bands can not hybridize along the $\Gamma$-Z line.     The topological property in AMgBi (A$=$K, Rb, Cs) can be  confirmed by analyzing its $Z_2$ topological invariance in the $k_z=0$ and $k_z=\pi$ planes. Due to the presence of the IS, we can analyze the $Z_2$ topological invariance in the view of point of the parities of the occupied bands in these two planes\cite{Z2 inversion}.   The parities of the $\Gamma_6$ and $\Gamma_7$ bands are shown to be even and odd, respectively.  The SOC causes the band inversion between the $\Gamma_7$ and $\Gamma_6$ bands only at the $\Gamma$ point. Thus, the Z$_2$ topological invariance in the $k_z=0$ plane is nontrivial while it is trivial in the $k_z=\pi$ plane. The topological property of AMgBi (A$=$K, Rb, Cs) can also be verified from their surface states and Fermi arcs, as shown in Ref.\cite{KMgBi}.

%
%

As the $\Gamma_6$ state is the anti-bonding state between two Bi atoms, the gap between the $\Gamma_6$ and $\Gamma_7$ bands is sensitive to the bond length between two Bi atoms. With the increase of the distance between two Bi atoms, the $\Gamma_6$ band shifts down in energy, which is helpful to topologically nontrivial phase. Tensile strain along $c$-axis is expected to be helpful for the band inversion. Substituting K atoms with bigger atoms will have the same effect. It clearly shows that larger lattice parameters are more in favor of the TDS phase, and the lattice parameter in $c$-direction has larger effect in Ref.\cite{KMgBi}. In-plane compressive strain has a similar effect, since the in-plane compressive strain always induces a tensile strain along $c$-axis\cite{KMgBi}.

%
%

\begin{figure}
\centerline{\includegraphics[width=0.45\textwidth]{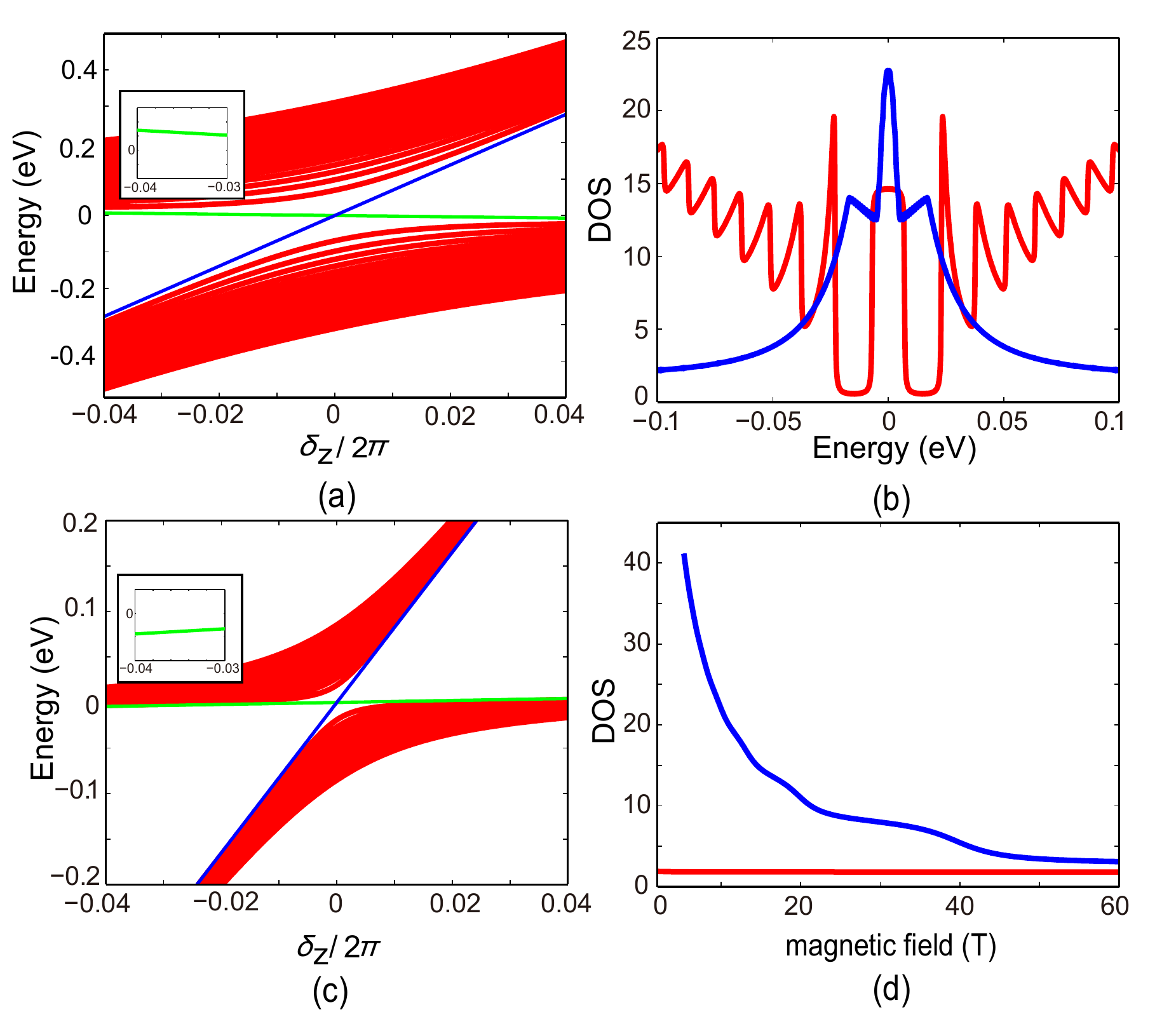}}
\caption{(color online) (a) and (c) are the Landau levels for KMgBi and RbMgBi with a magnetic field B=10T along the $z$-direction, respectively. (b) The DOSs for KMgBi and RbMgBi when the magnetic field is 10T; The DOSs of KMgBi is magnified by a factor of eight. (d) The  DOSs at the Fermi level  as a function of magnetic field. The red (blue) line stand for KMgBi (RbMgBi).
\label{AMgBilandau} }
\end{figure}

Compared to other TDSs\cite{Na3Bi,Cr3As2,PdTe1,PtTe}, AMgBi (A$=$K, Rb, Cs) are different. The $\Gamma_7$ band which contributes to the Dirac point has rather weak dispersion along the $z$-direction. The energy of the $\Gamma_7$ band at $\Gamma$ point for KMgBi is only about $3.8$ meV higher than that at Z point while the $\Gamma_7$ band at $\Gamma$ point is about 52 meV lower than that at Z point for RbMgBi, as shown in Fig.\ref{AMgBiband} (a) and (b), respectively. Therefore, they are both topological critical Dirac semimetals\cite{KMgBi}.  They both locate near the boundary of the type-\uppercase\expandafter{\romannumeral1} and type-\uppercase\expandafter{\romannumeral2} TDS phases. The band dispersion along the $\Gamma$-Z line also leads to that KMgBi is a type-\uppercase\expandafter{\romannumeral1} TDS while RbMgBi is a type-\uppercase\expandafter{\romannumeral2} TDS. This  is also supported by their FS conditions: when the Fermi level is tuned to be perfect at the Dirac points, the FSs of KMgBi consist of two discrete points at ($0, 0, \pm k_z^c$) while  those of RbMgBi contain an electron pocket and a hole pocket connecting with each other through the Dirac points, shown in Fig.\ref{AMgBicrystal}(b). As  shown in Ref.\cite{KMgBi},  a minimal four band model can be achieved to describe the low energy property of AMgBi near the $\Gamma$ point from the $k\cdot p$ method in the basis $|\Gamma_6^+,\frac{1}{2}\rangle, |\Gamma_7^-,\frac{3}{2}\rangle, |\Gamma_6^+,-\frac{1}{2}\rangle, |\Gamma_7^-,-\frac{3}{2}\rangle$

\begin{equation}
H_{0}({\bf k})  = \epsilon_0({\bf k})+ \left(\begin{array}{cccc}
M({\bf k})   &  ak_{-}  &     0    &  0      \\
ak_{+} &   -M({\bf k})  &     0    &  0      \\
0      &    0     &    M({\bf k})  & -ak_{+} \\
0      &    0     &  -ak_{-} & -M({\bf k})   \\
\end{array}\right),
\label{AMgBikp}
\end{equation}
where 
$\epsilon_0({\bf k})=C_0+C_1k^2_{z}+C_{2}(k^2_{x}+k^2_{y})$, $M({\bf k})=M_0-M_{1}k^2_{z}-M_{2}(k^2_{x}+k^2_{y})$ and $k_{\pm}=k_x{\pm}ik_y$. The parameters can be got by fitting the band structures\cite{KMgBi}. We mainly focus on the $C_1$ and $M_1$ parameters because the TDS phases can be directly determined by the two parameters: $|C_1|<|M_1|$ for type-\uppercase\expandafter{\romannumeral1} and $|C_1|>|M_1|$ for type-\uppercase\expandafter{\romannumeral2}. From our calculation,  we have $\frac{|C_1|}{|M_1|}=0.9505$ for KMgBi and $\frac{|C_1|}{|M_1|}=1.0268$ for RbMgBi. Therefore, KMgBi is a type-\uppercase\expandafter{\romannumeral1} TDS while RbMgBi is a type-\uppercase\expandafter{\romannumeral2} TDS, and both of them locate near the boundary of type-\uppercase\expandafter{\romannumeral1} and type-\uppercase\expandafter{\romannumeral1} TDS phases.

In the presence of magnetic field, TSs have many exotic transport properties\cite{Cr3As2 MR,BiSb MR,Narayanan2015,Feng2015,He2015,Novak2015}, which are usually closely related to the Landau levels contributed by the Weyl/Dirac points. Earlier studies\cite{Yao2016,Udagawa,Tchoumakov,Gu2011,Lukose2007} have shown that, the Landau levels for the type-\uppercase\expandafter{\romannumeral2} Dirac/Weyl semimetals collapse due to open quasi-classical orbitals when the direction of the magnetic field is out of the Dirac cone contributed by the Dirac/Weyl point\cite{type2}, while for type-\uppercase\expandafter{\romannumeral1} Dirac/Weyl semimetals the Landau levels are independent of the direction of the magnetic field. Therefore, when the magnetic field is applied perpendicular to the $z$-direction, the Landau levels of KMgBi are well-defined but close to collapse, but in RbMgBi they will collapse due to the type-II nature of the Dirac points. When the magnetic field is applied along the $z$-direction, the Landau levels always exist for both cases. For both KMgBi and RbMgBi, there is one nearly nondispersive chiral mode among the Landau levels, which is a reflection of the criticality of the bulk Dirac point, as shown in Fig.\ref{AMgBilandau} (a) and (c). The flat chiral mode leads to a large density of states (DOSs) near the Fermi level, shown in Fig.\ref{AMgBilandau}(b). However, there are key differences between the two cases. The two chiral modes of KMgBi have different velocities---one has a positive velocity and the other a negative velocity; while for RbMgBi, the velocity of both chiral modes is positive, shown in Fig.\ref{AMgBilandau}(a) and (c). This leads to the result that at large $k_z$, the $n\neq 0$ Landau level can always cross the Fermi level for RbMgBi, which is not the case for KMgBi. Hence, RbMgBi has a much higher DOSs at the Fermi level than KMgBi. For the fact that the $n=0$ Landau level is independent of the magnetic field, the DOSs at the Fermi level of KMgBi does not vary with the magnetic field. Nevertheless, since the energy gap between different Landau levels  increases with increasing magnetic field, the DOSs at the Fermi level of RbMgBi decreases rapidly, shown in Fig.\ref{AMgBilandau}(d).

Recently, a transport experiment has been carried out in KMgBi.  The measurement shows that  KMgBi is a narrow gap semiconductor with a small gap about 11.2 meV\cite{Zhang2016}, which is very close to topologically nontrivial phase as we discussed above. Based on this result and the data represented in Ref.\cite{KMgBi}, RbMgBi and CsMgBi should be in the topological critical Dirac semimetal phase.  With external in-plane compressive strain, KMgBi can also be tuned to be a topological critical Dirac semimetal\cite{KMgBi}.

\section{Topological materials in ternary compounds ABC (KZnP, BaAgAs, NaAuTe and KHgSb)}

\begin{figure}
\centerline{\includegraphics[width=0.45\textwidth]{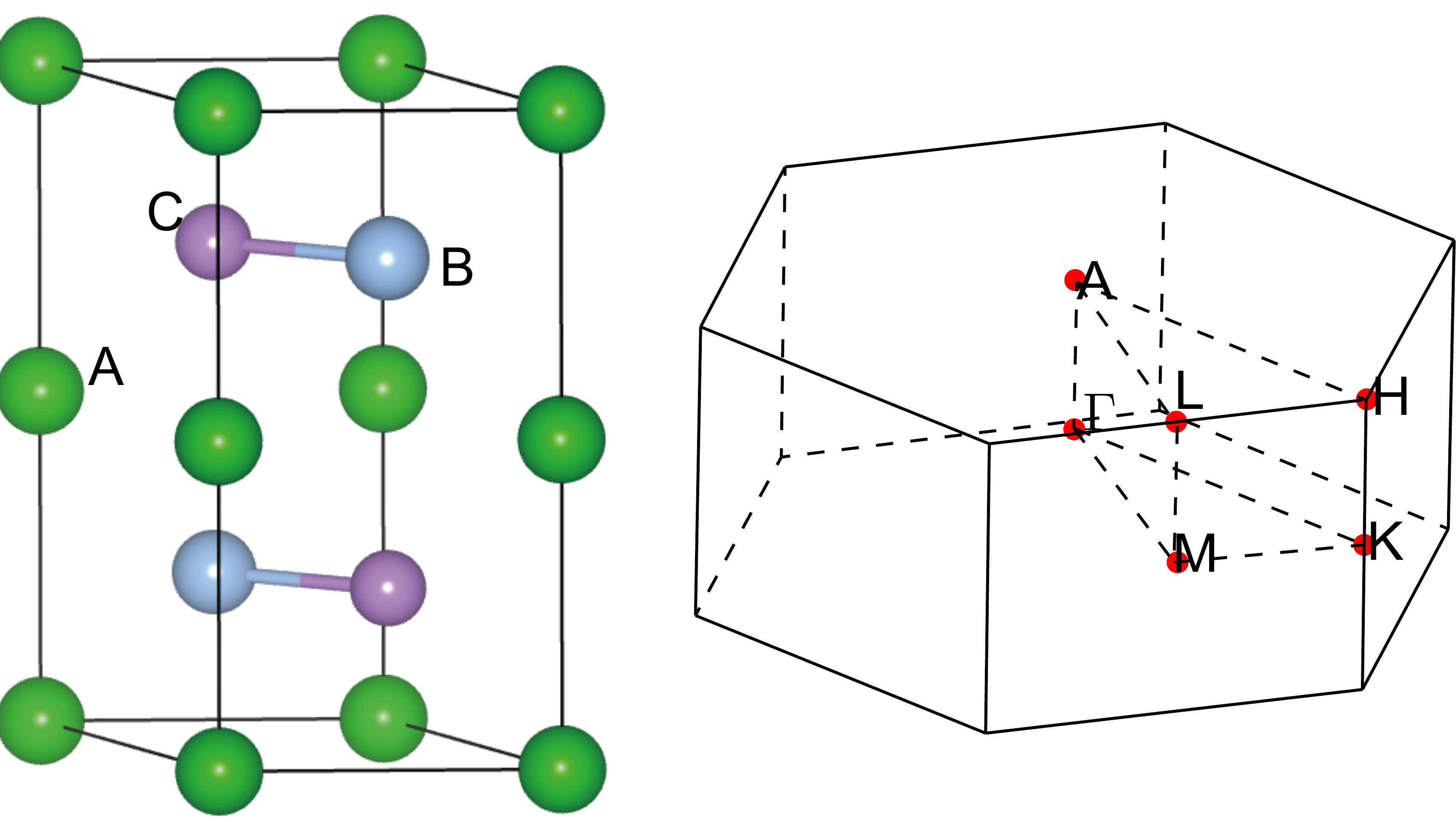}}
\caption{(color online) (a) Crystal structure for KZnP, BaAgAs, NaAuTe and KHgSb with the A atoms standing  for (K, Ba, Na), B for (Zn, Ag, Au, Hg), C for (P, As, Te, Sb). (b)  The high symmetry lines and points in the first BZ.
\label{NaAuTecrystal}}
\end{figure}

In this section, we discuss the topological property of a series of ternary compounds ABC (KZnP, BaAgAs, NaAuTe and KHgSb) whose space group is $P6_3/mmc$\cite{BaAgAs1,BaAgAs2,NaAuTe,hourglass}. The crystal structure of these compounds is similar to the hexagonal ZrBeSi type structure\cite{structure}, shown in Fig.\ref{NaAuTecrystal}(a). There are two BC layers in one unit cell, and the BC layers form triangular lattices and are sandwiched between trigonal $A$ layers along the $c$-axis. There are several symmetries in this crystal structure which play essential roles in the topological classification, including the mirror reflection $M_z$ perpendicular to the $c$-axis, the rotational symmetry and rotation-translational symmetry along the $c$-axis, and the glide plane symmetry along the $c$-axis. The realization of various topological phases in these compounds depends on the competition between several energy scales: the energy of the atomic orbitals, the energy of the crystal splitting, the energy difference between the bonding and antibonding states, and the strength of the SOC.

\begin{figure}
\centerline{\includegraphics[width=0.45\textwidth]{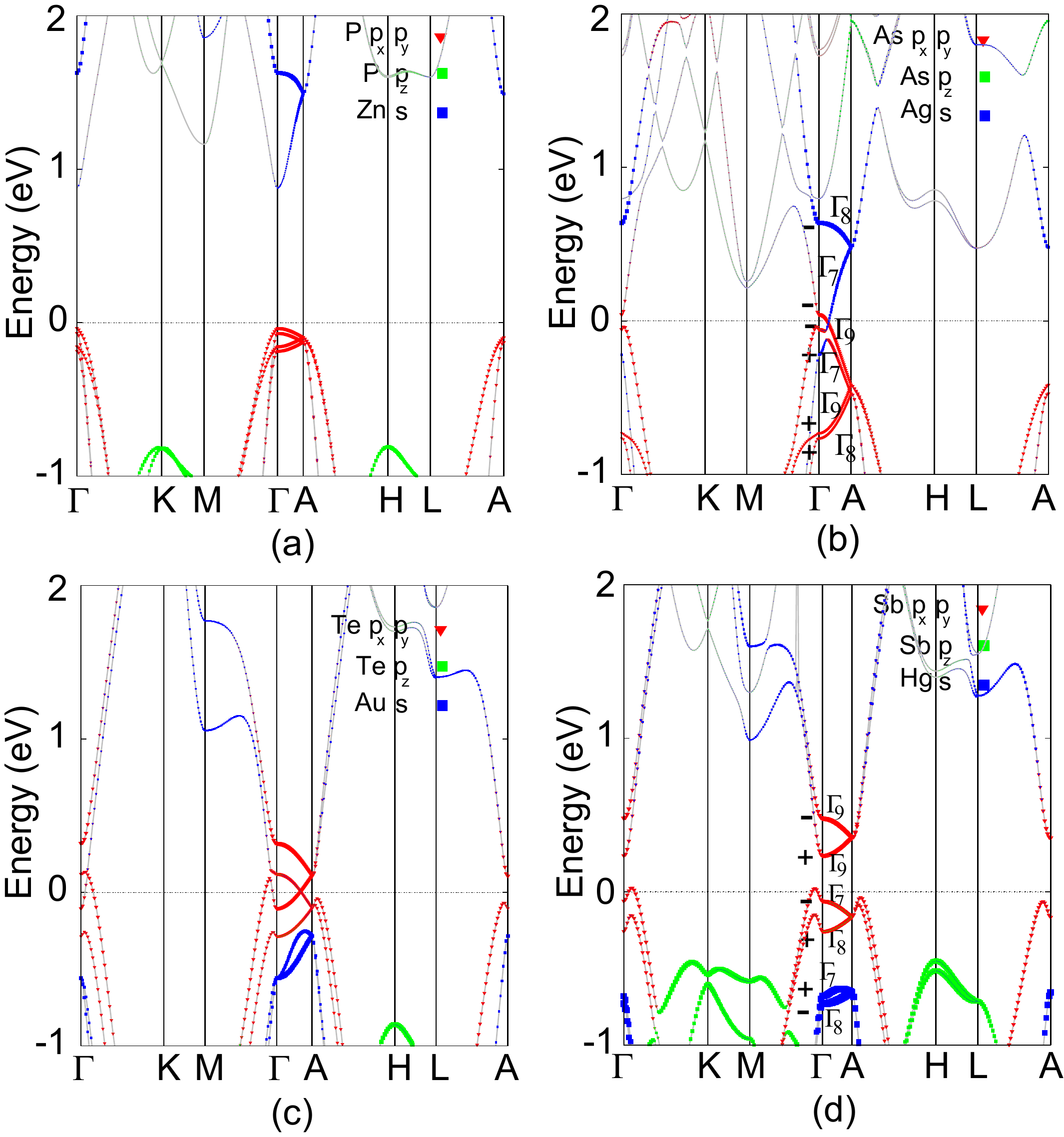}}
\caption{(color online)The band structures for KZnP in (a), BaAgAs in (b), NaAuTe in (c) and KHgSb in (d), respectively. The parity and irreducible representation for the bands near the Fermi level are labeled by $\pm$ and $\Gamma_i$, respectively.
\label{NaAuTeband}}
\end{figure}

\begin{figure}
\centerline{\includegraphics[width=0.45\textwidth]{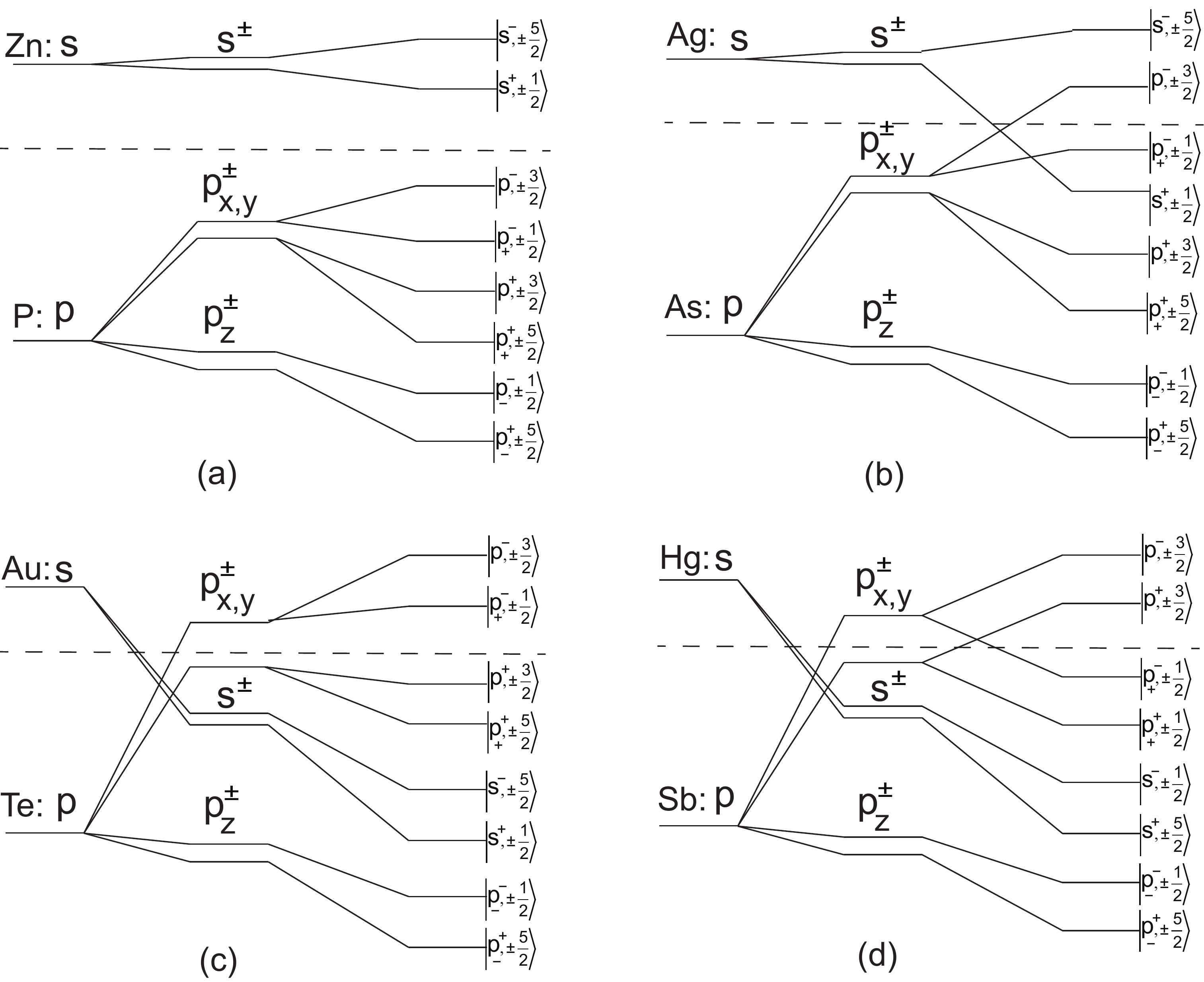}}
\caption{(color online) The energy split for the $s$-orbital and $p$-orbital near the Fermi level with SOC at the $\Gamma$ point for KZnP in (a), BaAgAs in (b), NaAuTe in (c) and KHgSb in (d), respectively.
\label{NaAuTesplit}}
\end{figure}

We first analyze the energy scales. As   shown in Fig.\ref{NaAuTeband}, the gap minima appears along the $\Gamma$-A line and the states near the Fermi level are mainly contributed by the $s$-orbital of the B atoms and the $p$-orbital of the C atoms.  We can focus on  the energy split of these orbitals. In the atomic limit, the energy of the $s$-orbital of the B atoms tends to be higher than that of the $p$-orbital of the C atoms, because the principal quantum number of the $s$-orbital is larger than that of the $p$-orbital, as shown in Table.\ref{NaAuTeonsiteenergy}. The second energy scale is the energy of the crystal splitting. Both the B atoms and C atoms in the compounds are in a triangle environment, and the in-plane distance between the B and C atoms is much smaller than that along the $c$-axis, as shown in Table.\ref{NaAuTeparameter}. However, Considering B being cations and C being anions,   the energy of the $p_z$-orbital is  higher than that of the $p_{x,y}$-orbital, as shown in Table.\ref{NaAuTeonsiteenergy}. Since the inversion center locates between the two BC layers in a unit cell, the orbitals from the two sublayers form bonding and antibonding states
\begin{eqnarray}\label{bounding}
|s,+\rangle &=& \frac{1}{\sqrt{2}}(|s_a\rangle+|s_b\rangle), \nonumber\\
|s,-\rangle &=& \frac{1}{\sqrt{2}}(|s_a\rangle-|s_b\rangle), \nonumber\\
|p,+\rangle &=& \frac{1}{\sqrt{2}}(|p_a\rangle-|p_b\rangle), \nonumber\\
|p,-\rangle &=& \frac{1}{\sqrt{2}}(|p_a\rangle+|p_b\rangle),
\end{eqnarray}
where $'+'$ ($'-'$) in the kets stands for the bonding (antibonding) state, and $a$ and $b$ are the sublayer indices. The energy of the antibonding state is always higher than that of the bonding state, and the parity of the bonding (antibonding) state is always even (odd).  The energy difference between the bonding and antibonding states is mainly affected by the lattice parameter along the $c$-axis. The last energy scale is the strength of the SOC in these compounds. Both the strength of the atomic SOC and the environment of the ions can affect the effective SOC. Since all of the compounds considered here have the same crystal structure, the effective SOC in these compounds is mainly affected by the strength of the atomic SOC. In general,  heavier ions have larger atomic SOC. When the SOC varies, the states with larger magnetic quantum number $J_z$   have higher energy. Fig.\ref{NaAuTesplit} shows these energy splittings for KZnP, BaAgAs, NaAuTe and KHgSb, respectively.

To analyze the topological phases in these compounds, we start with KHgSb, which has been verified to be the first topological hourglass insulator\cite{hourglass,hourglass experiment}. As shown in Fig.\ref{NaAuTeband}(d), at $\Gamma$ point the bands near the Fermi level are $|\Gamma_9,\pm\rangle=|\frac{3}{2},\pm\rangle$, $|\Gamma_7,\pm\rangle=|\frac{1}{2},\pm\rangle$ and $|\Gamma_8,\pm\rangle = |\frac{5}{2},\pm\rangle$. Since there are IS and TRS in these compounds, all of the states are doubly degenerate, but we just take one state of each Kramers' pair in our analysis for simplicity. The $\Gamma_9$ states are contributed by the $p_{x,y}$-orbital, and the $\Gamma_7$ and $\Gamma_8$ states are mainly contributed by the $p_{x,y}$-orbital and $s$-orbital. It is interesting that the $p_{x,y}$-orbital and $s$-orbital can form the $J_z=\frac{5}{2}$ states, which is impossible for the atomic orbitals. It is the nonsymmorphic symmetry which takes the responsibility for the $J_z=\frac{5}{2}$ states. For instance, Eq.\ref{bounding} shows that, the $s$-orbital can form bonding and antibonding states $|s,+\rangle$ and $|s,-\rangle$. When SOC is turned on, along the $\Gamma$-A line, the four states which are the eigenstates of the rotational symmetry along the $c$-axis can be expressed as

\begin{eqnarray}\label{boundingJz}
|s,J_z^1,+\rangle &=& \frac{1}{\sqrt{2}}(|s_a,\uparrow\rangle+e^{i\frac{k_z}{2}}|s_b,\uparrow\rangle), \nonumber\\
|s,J_z^2,+\rangle &=& \frac{1}{\sqrt{2}}(|s_a,\downarrow\rangle+e^{i\frac{k_z}{2}}|s_b,\downarrow\rangle), \nonumber\\
|s,J_z^3,-\rangle &=& \frac{1}{\sqrt{2}}(|s_a,\uparrow\rangle-e^{i\frac{k_z}{2}}|s_b,\uparrow\rangle), \nonumber\\
|s,J_z^4,-\rangle &=& \frac{1}{\sqrt{2}}(|s_a,\downarrow\rangle-e^{i\frac{k_z}{2}}|s_b,\downarrow\rangle),
\end{eqnarray}
where the $k_z$ dependance in the kets has been omitted for simplicity, and   the phase factor on the right side stems from the fact that the distance between the two sublayers is $\frac{1}{2}c$ along the $c$-direction, where $c$ is the lattice parameter along the $c$-axis. We  classify the states in Eq.\ref{boundingJz} according to the mirror symmetry $M_z$ as  following. The mirror symmetry $M_z$ can be viewed as the combination of the IS and the $C_{2z}$ rotational symmetry. The eigenstates of the $C_{2z}$ rotational symmetry are usually the eigenstates of the mirror symmetry $M_z$. However, the condition is more complicated for the space group $P6_3/mmc$ because of its nonsymmorphic property. In this space group, the $C_{2z}$ ($C_{6z}$) rotational symmetry is not well defined, but the rotation-translational symmetry $\{C_{2z}|\frac{1}{2}c\}$ ($\{C_{6z}|\frac{1}{2}c\}$) is well defined, where $\{R|t\}$ means a translation $t$ after a point group operation $R$. After some algebra, the commutation relation between the symmetry $\{C_{2z}|\frac{1}{2}c\}$ and $M_z$ can be achieved as $[\{C_{2z}|\frac{1}{2}c\},M_z]=c$. In the reciprocal space, this is $[\{C_{2z}|\frac{1}{2}c\},M_z]=e^{ik_z}$ at $\Gamma$ and A points.

At $\Gamma$ point, the nonsymmorphic operation equals to a point group operation, since the fractional translation in the nonsymmorphic operation is futile when $k_z=0$. The little group for KHgSb at $\Gamma$ is $D_{6h}$.  However, the $C_6$ rotation switches the two sublayers in a unit cell, which is different from the simple space groups. Thus,  when $C_6$ acts on the states in Eq.\ref{boundingJz}, besides the phase $e^{\pm i\frac{\pi}{6}}$, there  is  another $'+'$ ($'-'$) for the even (odd) parity states. For the odd parity states, the $'-'$ phase  changes $J_z$ to $J_z-3$ while for the even parity states there is no difference. For the states contributed by $p$-orbital, the consequence is inversed for the odd parity of the $p$-orbital. Based on the above analysis, we can get the $J_z^i$ in Eq.\ref{boundingJz} as $J_z^1=-J_z^2=\frac{1}{2}$ and $J_z^4=-J_z^3=\frac{5}{2}$ at $\Gamma$.  The analysis is invalid at A point because the fractional translation in the nonsymmorphic symmetry has effect on the classification of states\cite{Zak,nonsymmorphic}: the $\{C_{2z}|\frac{1}{2}c\}$ symmetry and other related symmetries anticommute with the mirror symmetry $M_z$.

KZnP is simply a NI. However, for KHgSb,  there are band inversions both at $\Gamma$ and A points between the $|\Gamma_7,+\rangle$, $|\Gamma_8,-\rangle$ and $|\Gamma_9,+\rangle$, $|\Gamma_9,-\rangle$ states as shown in Fig.\ref{NaAuTeband}(a)(d) and Fig.\ref{NaAuTesplit}(a)(d).  The band inversions stem from the on-site energy difference between the $s$-orbital and $p_{x,y}$-orbital, which is much smaller than those in  KZnP, as shown in table.\ref{NaAuTeonsiteenergy}. The total parity for the occupied states remains unchanged.  Therefore, KHgSb is still a NI according to the $Z_2$ classification based on the TRS\cite{Z2 inversion}. However, there are mirror symmetries in KHgSb and the band inversions give rise to nonzero mirror Chern numbers based on $M_z$. The mirror Chern numbers are considered within the two mirror invariant planes $k_z=0,\pi$. In the $k_z=0$ plane, the inverted bands can be classified according to their eigenvalues of $M_z$: $|\Gamma_7,\frac{1}{2},+\rangle$, $|\Gamma_8,-\frac{5}{2},-\rangle$, $|\Gamma_9,-\frac{3}{2},+\rangle$ and $|\Gamma_9,\frac{3}{2},-\rangle$ belong to the $+i$ subspace.  Their Kramers' pairs belong to the $-i$ subspace. In the $k_z=0$ plane, since the band inversion occurs only at $\Gamma$ point, the mirror Chern number $M_0$ can be obtained directly by calculating the eigenvalues of the inverted bands in each mirror subspace based on the $D_{6h}$ point group\cite{Chern number group,mirror Chern group}
\begin{eqnarray}\label{mirrorChern}
e^{i\frac{\pi}{3}M_0} = e^{i\frac{\pi}{3}(J_{z,|\frac{1}{2},+\rangle}+J_{z,|-\frac{5}{2},-\rangle}-J_{z,|-\frac{3}{2},+\rangle}-J_{z,|\frac{3}{2},-\rangle})},
\end{eqnarray}
where $M_0=-2$. The shift of $J_z$ to $J_z-3$, which is caused by the nonsymmorphic symmetry, has played an essential role in the realization of the topological nontrivial mirror Chern number. In the $k_z=\pi$ plane, the mirror Chern number is always zero, which is protected by symmetries, as shown in ref.\cite{hourglass}.


\begin{table}[bt]
\caption{\label{NaAuTeparameter}%
Optimized structural parameters
for KZnP, BaAgAs, NaAuTe and KHgSb by using GGA in the paramagnetic phase.}

\begin{ruledtabular}
\begin{tabular}{cccc}
  & a(\AA) & c(\AA) & B-C(\AA)   \\
 \colrule
 KZnP  &  4.106 & 10.304 & 2.371 \\
BaAgAs & 4.670  & 9.119  & 2.696 \\
NaAuTe & 4.626  & 8.126  & 2.671 \\
KHgSb  &4.897   & 10.523 & 2.827 \\
\end{tabular}
\end{ruledtabular}
\end{table}

\begin{table}[bt]
\caption{\label{NaAuTeonsiteenergy}%
The on-site energy of the $s$-orbital for Zn, Ag, Hg, Au, and $p$-orbitals for P, As, Sb, Te, in the unit of eV.}

\begin{ruledtabular}
\begin{tabular}{ccccc}
           & KZnP     & BaAgAs   & NaAuTe  & KHgSb \\
 \colrule
 p$_{x,y}$ & -0.2906  & 1.1967   & 1.6631  & 1.2248 \\
 p$_z$     & 0.9998   & 1.7909   & 2.2841  & 1.5348 \\
 s         & 3.0532   & 3.7792   & 4.0193  & 1.2924 \\
\end{tabular}
\end{ruledtabular}
\end{table}

NaAuTe is a TDS, which is quite similar to BaAuBi\cite{NaAuTe}. Compared to KHgSb, there are two obvious differences. The $|\Gamma_7,+\rangle$ and $|\Gamma_8,-\rangle$ bands contributed by the $s$-orbital of the Au atoms are closer to the Fermi level, as shown in Fig.\ref{NaAuTeband}(c). This is due to the fact that the difference of the on-site energy between the Au $s$-orbital and the Te $p$-orbital is much larger than that in KHgSb as  shown in Table,\ref{NaAuTeonsiteenergy}.
The second difference is that, the $|\Gamma_7,-\rangle$ and $|\Gamma_9,+\rangle$ bands contributed by the Te $p_{x,y}$-orbital cross at $(0, 0, \pm k_z^c)$ to result in two Dirac points near the Fermi level as shown in Fig.\ref{NaAuTeband}(c). Since the strength of the SOC for Te and Sb are similar, the band crossings stem from the fact that the energy difference between the bonding and antibonding states in NaAuTe is much larger than that in KHgSb shown in Fig.\ref{NaAuTeband}(c) and Fig.\ref{NaAuTesplit}(c). It is also shown in Table.\ref{NaAuTeparameter}: the lattice parameter in $c$-axis for NaAuTe is much smaller than that for KHgSb, so  that the bonding  between the two AuTe layers in a unit cell is much stronger than that in KHgSb.  The TDS phase in NaAuTe type materials is rather robust. Only when the energy scale of SOC is larger than the energy difference between the bonding and antibonding states, the two Dirac points will gap out and the system becomes a KHgSb type insulator.   Furthermore, we can consider the mirror Chern number in NaAuTe. Similar to KHgSb, the mirror Chern number in $k_z=\pi$ plane is $0$. In $k_z=0$ plane, the mirror Chern number is $M_0=-3$.  Compared with KHgSb,  there is another band inversion between the $|\Gamma_7,-\rangle$ and $|\Gamma_9,+\rangle$ bands. Based on Eq.\ref{mirrorChern}, we can show that this band inversion contributes another $-1$ to the mirror Chern number $M_0$. The $Z_2$ invariant  is nontrivial in the $k_z=0$ plane but  trivial in the $k_z=\pi$ plane.

BaAgAs is also a TDS\cite{BaAgAs2}.  However, different from NaAuTe, there are band crossings in BaAgAs between the $|\Gamma_7,+\rangle$ and $|\Gamma_9,-\rangle$ bands, as shown in Fig.\ref{NaAuTeband}(b). The $|\Gamma_7,+\rangle$ band has main contribution of the Ag $s$-orbital and the $|\Gamma_9,-\rangle$ band is mainly contributed by the As $p_{x,y}$-orbital.   BaAgAs has much weaker SOC than NaAuTe and KHgSb.  The two dominated energy scales are the energy scale of bonding and antibonding states and the energy scale of the on-site energy of the $s$-orbital and $p$-orbital, as shown in Fig.\ref{NaAuTeband}(b) and Fig.\ref{NaAuTesplit}(b). These two energy scales are  between that of KZnP and NaAuTe, shown in Table.\ref{NaAuTeparameter} and Table.\ref{NaAuTeonsiteenergy}. As a result, the $|\Gamma_7,+\rangle$ band in BaAgAs has higher energy than that in NaAuTe. It  crosses with the $|\Gamma_9,-\rangle$ band, leading to the TDS phase. Similarly, the mirror Chern number can be calculated in the $k_z=0$ plane to be $-1$, and the $Z_2$ invariant in the $k_z=0$ plane is nontrivial while it is trivial in the $k_z=\pi$ plane.

The above results show that various topological phases can be realized in the series of ternary compounds ABC (KZnP, BaAgAs, NaAuTe and KHgSb).  Moreover, the topological phases are sensitive to lattice parameters and the change of atoms. Therefore,  different topological phase transition can be realized by element substitution.

\section{summary and outlook}

In summary, we have shown that the series of materials AMgBi (A=K, Rb, Cs) are  topological critical Dirac semimetals. They can be a very special playground to explore semimetal physics that is absent in the previously known DSs Na$_3$Bi and Cd$_3$As$_2$.  Flat bands and flat Landau levels can be formed in the vicinity of  the boundary between type-\uppercase\expandafter{\romannumeral1} and type-\uppercase\expandafter{\romannumeral2} Dirac semimetal phases. Thus, the effect of electron-electron correlation in these materials can be strong. It will be interesting to explore the intriguing many-body emergent physics.  Moreover, the type-\uppercase\expandafter{\romannumeral1} and type-\uppercase\expandafter{\romannumeral2} TDS phases can be easily adjusted by   strain and element substitution. The collapse of the Landau levels is expected to accompany the phase transition between the two phases.

The series of ternary compounds ABC (KZnP, BaAgAs, NaAuTe and KHgSb) can realize various topological insulating states and semimetal states. The realization of these topological states depends on the competition between several energy scales. Because of their unique crystal structure, their surface states can exhibit exotic electronic structures\cite{AA}.   The topological phases can also be effectively adjusted by element substitution. The topological mirror Chern numbers are also tunable.The TDS phase in NaAuTe is more robust than other TDSs due to its unique band structures. Furthermore, since NaAuTe has high mirror Chern number in its TDS phase, topological insulating states with high Chern number or high mirror Chern number can also be realized.

\section{acknowledgement}
This work is supported by the Ministry of Science and Technology of China 973 program(Grant No. 2015CB921300, No.2017YFA0303100), National Science Foundation of China (Grant No. NSFC-11334012), and   the Strategic Priority Research Program of  CAS (Grant No. XDB07000000).

\end{document}